\documentclass{PoS}

\title{Precision measurements of rare kaon decays}

\ShortTitle{Precision measurements of rare kaon decays}

\author{\speaker{Evelina MARINOVA}\thanks{On behalf of NA48/2 collaboration}\\
        INFN Sezione di Perugia\\
        E-mail: \email{Evelina.Marinova@cern.ch}}

\abstract{

We report the recent results on rare kaon decays from the NA48/2
experiment. The precise measurement  of direct photon emission (DE)
in the decay
 $K^\pm\rightarrow \pi^\pm \pi^0\gamma$ and its interference (INT), with the INT amplitude
 being observed for the first time, has been finalized. This study is based
 on the full NA48/2 data set with about 600k reconstructed $K^\pm \rightarrow \pi^\pm \pi^0
 \gamma$
 decays which is factor of 30 larger than for previous experiments.
Samples of about 7200 reconstructed $K^\pm \rightarrow \pi^\pm e^+
e^-$, and more than 3000 $K^\pm \rightarrow \pi^\pm \mu^+ \mu^-$
events, with very small background contamination, have been
collected. The latter is exceeding the total existing statistics by
a factor of five. A precise measurement of the branching fractions
and the form factors of the rare decays $K^\pm\rightarrow \pi^\pm
e^+ e^-$ has been performed using different theoretical models. The
CP violating asymmetry between $K^+$ and $K^-$ in this channel is
also measured.
 }

\FullConference{European Physical Society Europhysics Conference on High Energy Physics,
EPS-HEP 2009,\\
         July 16 - 22 2009\\
         Krakow, Poland}

\begin{document}

The NA48/2 experiment at the CERN-SPS, designed primarily for charge
asymmetry measurements~\cite{k3pi}, uses simultaneous $K^+$ and
$K^-$ beams and has collected large samples of rare decays. Detailed
information about the main detector and its performance can be found
elsewhere~\cite{na48det}.

\subsection*{$K^\pm \rightarrow \pi^\pm \pi^0
 \gamma$
 decays}
\begin{figure}[h!]
  \vspace{0.0pt}
  \centerline{
    \put(120.50,85.5){\small{(a)}}
    \put(254.50,85.5){\small{(b)}}
    \put(394.50,85.5){\small{(c)}}
    \includegraphics[width=5.10cm]{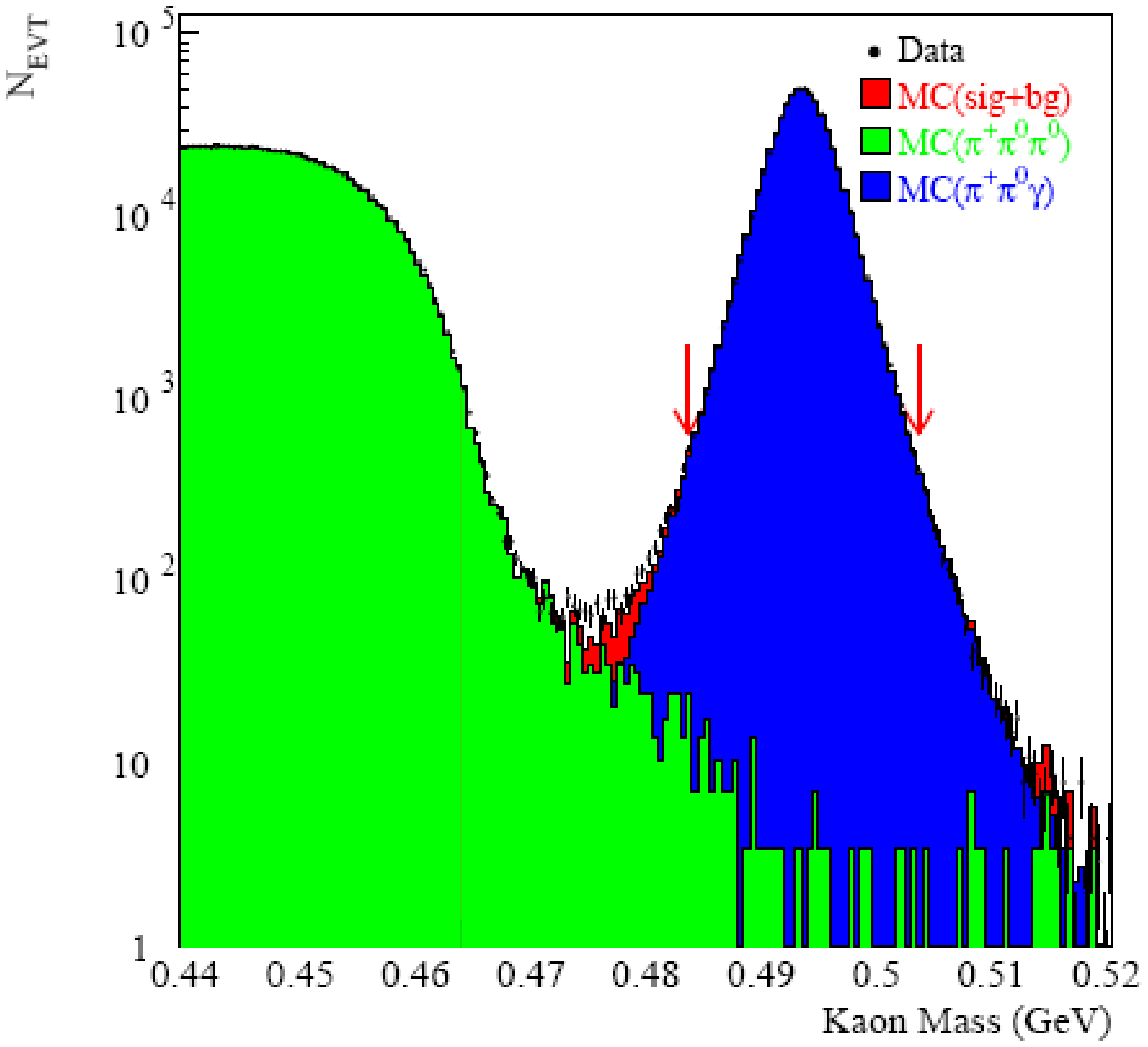}
    \includegraphics[width=4.35cm]{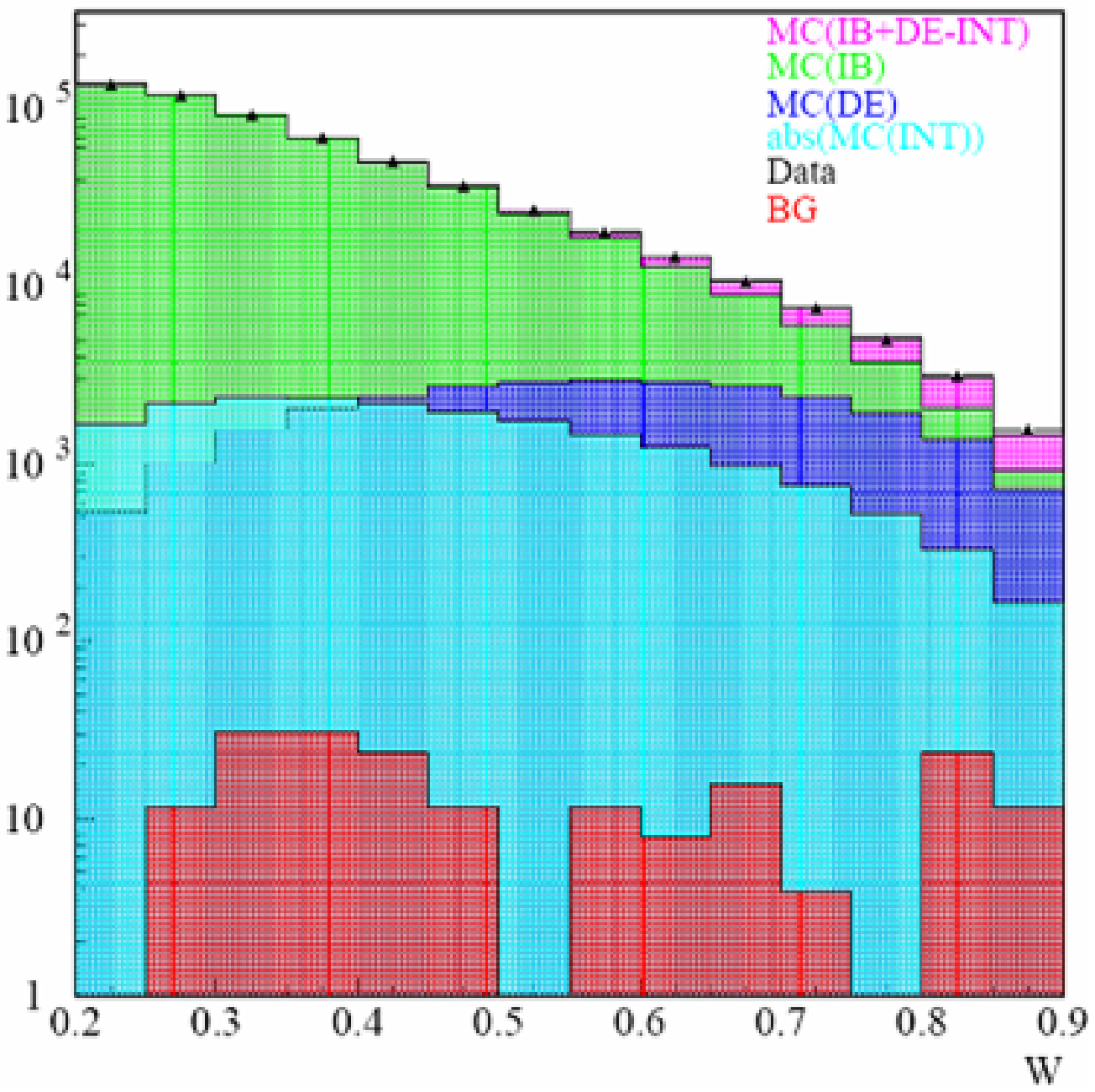}
    \includegraphics[width=4.95cm]{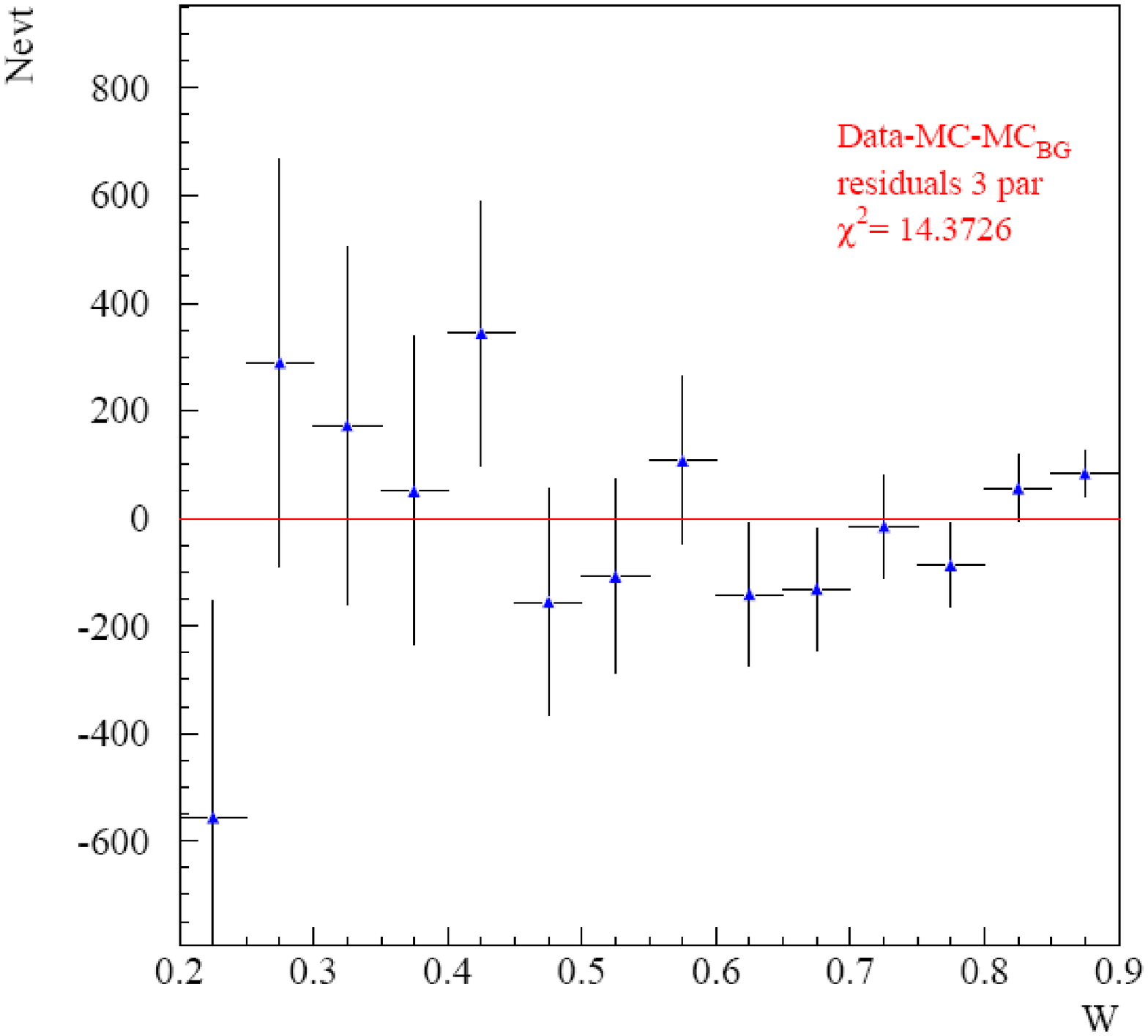}
     }\caption{(a) Reconstructed mass of the kaon; (b) Contribution of the different
 components to the decay rate,their sum and the data distribution; (c) The fit residuals distribution. \label{fig:pp0g}}
\end{figure}

The properties of the $K^\pm \rightarrow \pi^\pm \pi^0
 \gamma$ decays are described in terms of the Lorentz invariant $W$
 which is given by
 $W^2=(P_K^*\cdot P_{\gamma}^*)(P_{\pi}^*\cdot P_{\gamma}^*)/(m_K\cdot m_{\pi})^2$~\cite{ppg2,ppg3}, where $P_K^*,~ P_{\gamma}^*,~P_{\pi}^*$ are the particles momenta in the kaon rest frame.
Both Inner bremsstrahlung (IB) and Direct emission (DE) contribute
to the amplitude of the decay. DE can occur either through electric
or through magnetic dipole transition. The electric dipole
transition can interfere with the IB term giving rise to an
interference term with CP violating contributions. The previous
measurements of IB and DE, reported in reference~\cite{pdg}, are
performed in the region 55 to 90 MeV in terms of kinetic energy of
the pion in the kaon rest frame, $T^*_{\pi}$. The decay rate is
written as
\begin{equation}
\frac{d\Gamma^\pm}{dW}=\frac{d\Gamma_{IB}^\pm}{dW}\left[1 +
2\cos(\pm
\phi+\delta_1^1-\delta_0^2)m_{\pi}^2m_K^2X_EW^2+m_{\pi}^4m_K^4(|X_E|^2+|X_M|^2)W^4\right].
\end{equation}
The NA48/2 measurement is made in an enlarged region of $T^*_{\pi}$,
(0 - 80 MeV). After applying a cut on $W$ (0.2 - 0.9), about 600 000
events have been reconstructed, with a background estimation of
about 1\% of DE component (see Fig.~\ref{fig:pp0g} (a)). The
mistagging probability for the radiated gamma is as low as one per
mille. As the different contributions have a different W dependence,
IB, INT and DE distributions can be well separated. The main
technique used to extract INT and DE contributions is based on
Poissonian Maximum Likelihood fit. The relative contribution of
every source is calculated in such a way that the difference between
the number of events and the total number of simulated events in
every bin is minimal (see Fig.~\ref{fig:pp0g} (b), (c)). A
cross-check was performed using a polynomial fit.

The final results for the relative fraction of DE and INT with
respect to IB are $Frac(DE)_{T^*_{\pi}(0-80)~\mathrm{MeV}}= (3.32
\pm 0.15_{stat} \pm 0.14_{sys})\cdot 10^{-2}$,
$Frac(INT)_{T^*_{\pi}(0-80)~\mathrm{MeV}}=(-2.35 \pm 0.35_{stat} \pm
0.39_{sys})\cdot 10^{-2}$. In order to compare the result for DE,
INT was assumed to be 0, a polynomial fit was used, and $T^*_{\pi}$
is in the range (55 - 90) MeV. The obtained result $BR(DE)_{
T^*_{\pi}(55-90)MeV}= (2.3\pm0.05_{stat}\pm 0.077_{sys})\cdot
10^{-6}$ disagrees with the previous measurements~\cite{pdg} but the
bad $\chi^2$ of the fit indicates that $INT = 0$ is a wrong
assumption. The simultaneous measurement of INT and DE allows a
measurement of the electric and the magnetic contributions to DE.
The results are $X_E=(-24 \pm 4_{stat} \pm 4_{sys})
~\mathrm{GeV}^{-4}$, $X_M = (254 \pm 11_{stat} \pm
11_{sys})~\mathrm{GeV}^{-4}$.

 \subsection*{$K^\pm
\rightarrow \pi^\pm e^+ e^-$ decays}

\begin{figure}[h!]
  \vspace{0.5 pt}
  \centerline{
    \put(109.0,106.5){\small{(a)}}
    \put(241.50,106.5){\small{(b)}}
    \put(371.50,106.5){\small{(c)}}
    \includegraphics[width=4.5cm]{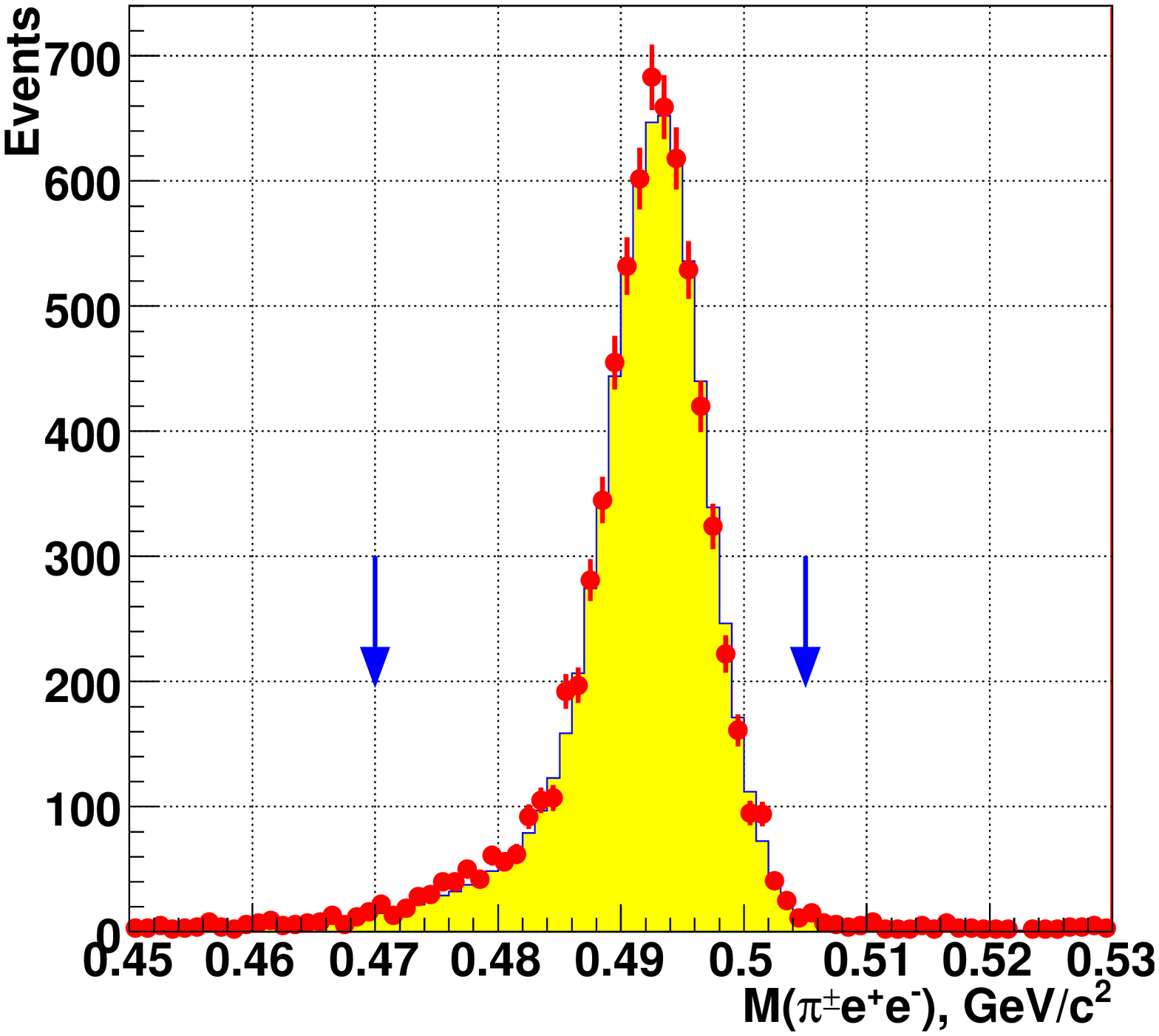}
    \includegraphics[width=4.53cm]{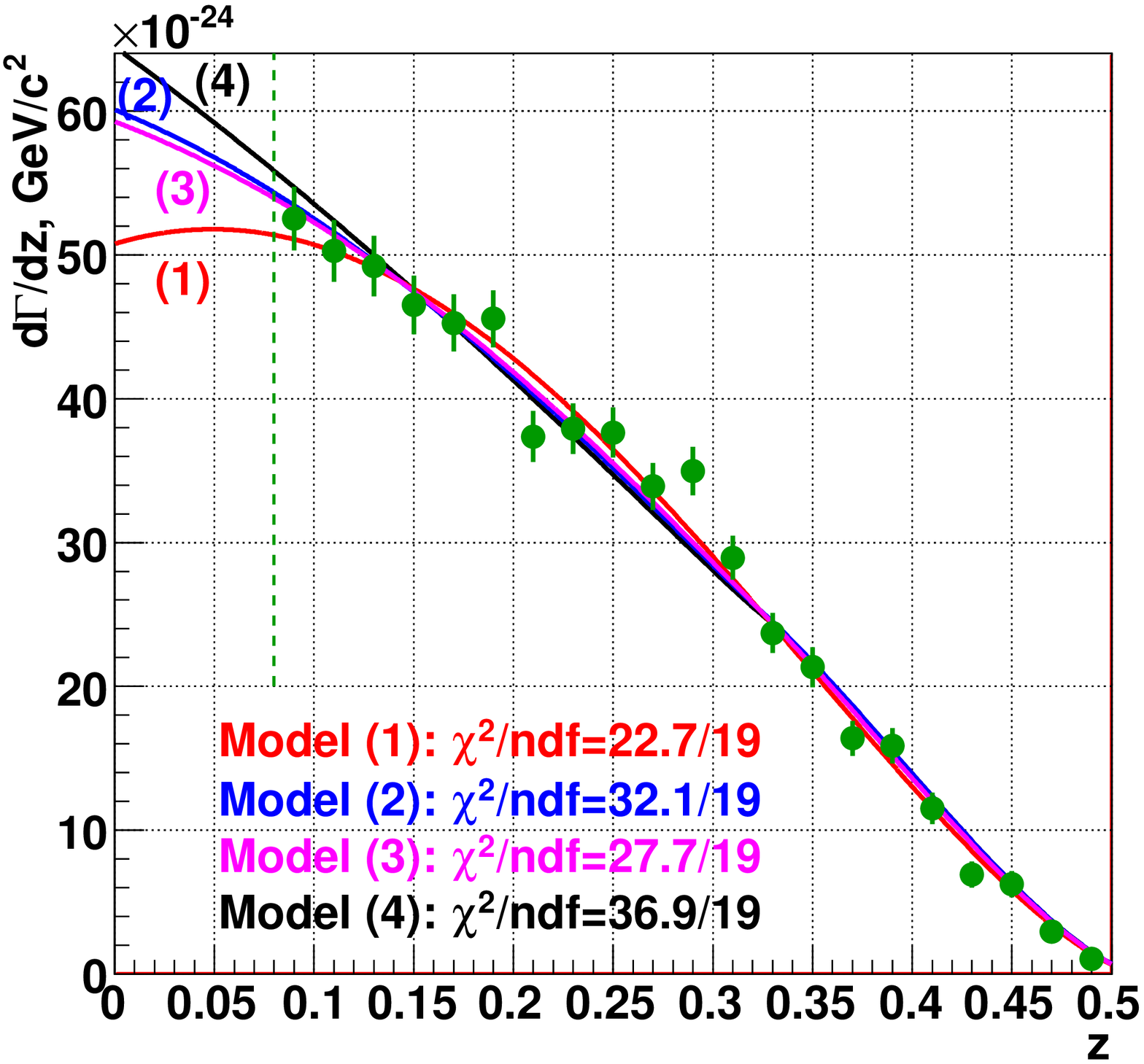}
    \includegraphics[width=4.5cm]{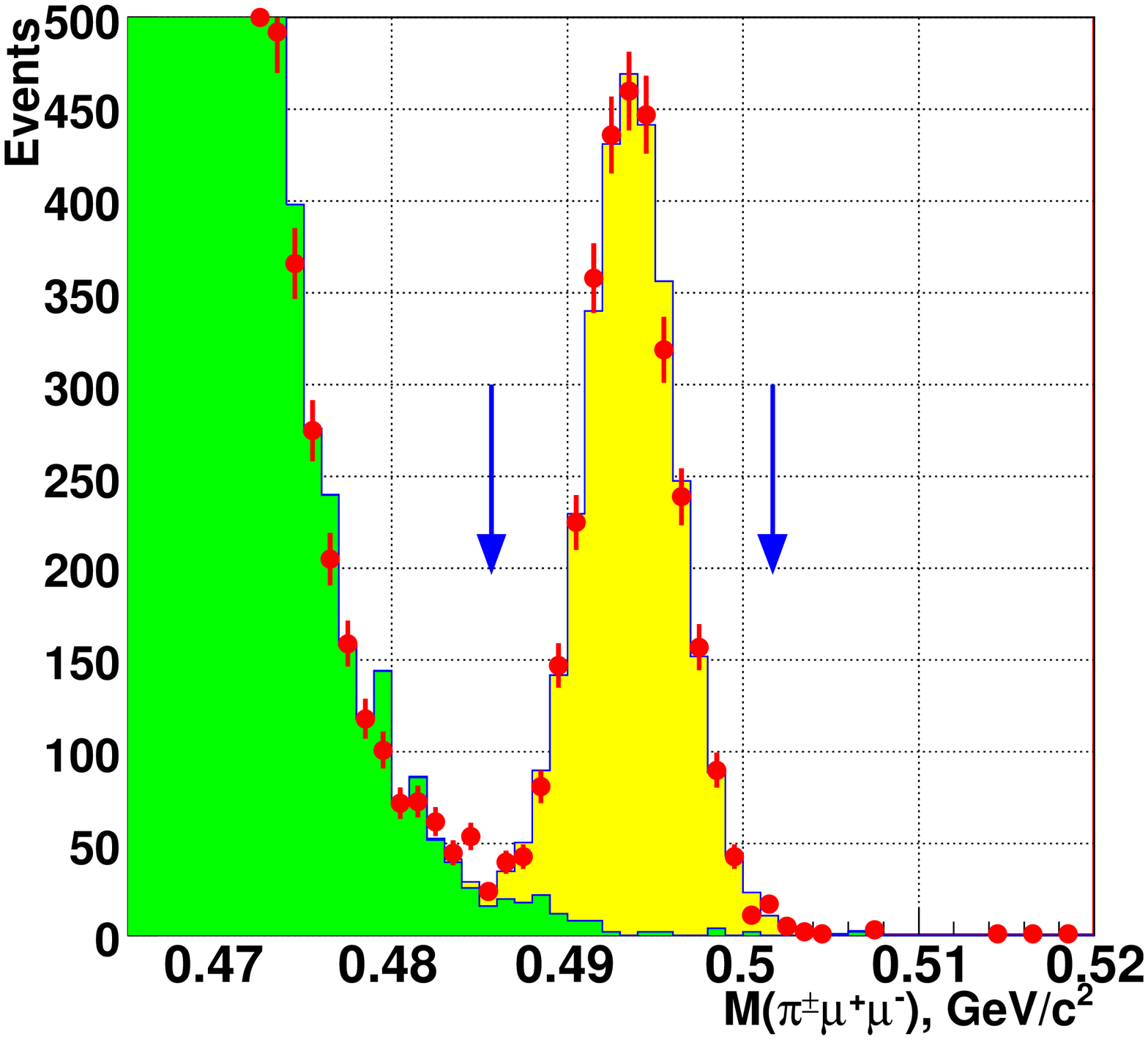}
    }\caption{(a) The invariant mass of the kaon, reconstructed from $K^\pm
\rightarrow \pi^\pm e^+ e^-$ decays;
 (b) The computed $(d\Gamma/dz)_i$
and the results of the $\chi^2$ of the fits corresponding to the
considered models; (c) The invariant mass of the kaon, reconstructed
from $K^\pm \rightarrow \pi^\pm \mu^+ \mu^-$ decays.
\label{fig:3fits}}
\end{figure}

The properties of the rare decays $K^\pm \rightarrow \pi^\pm \ell^+
\ell^-$ involving FCNC processes are described in terms of
$z=M^2_{\ell\ell}/M^2_K$. The decay rate is given by $d\Gamma/dz
\sim \rho(z) W(z)$ where $\rho(z)$ is a phase space factor. There
are four theoretical models predicting the form factors $W(z)$ (see
Tab.~\ref{tab:resultsrare}). Each model is characterized by two free
parameters which determine a model dependent branching fraction. The
measurements are done normalising to the
$K^\pm\rightarrow\pi^\pm\pi^0_{Dalitz}$ which has the same charged
particles in the final state. Due to the similarities between the
signal and normalisation channel, systematics due to particle
identification, trigger efficiencies, etc., is kept very low. In
total 7253 events have been reconstructed with a background
estimation of 1\% (see Fig.~\ref{fig:3fits}(a)). The kinematical
region in $z$ is restricted to be larger than 0.8 (corresponding to
$ M_{ee}>140$ MeV/c$^2$) in order to suppress the background from
the abundant normalisation channel. Fits to all models have been
performed and have a reasonable quality (see
Fig.~\ref{fig:3fits}(b).) The results are reported in
Tab.~\ref{tab:resultsrare}. The combined result for a model
dependent branching fraction is $BR =
(3.11\pm0.04_{stat}\pm0.05_{syst}\pm0.08_{ext}\pm 0.07_{model})\cdot
10^{-7}$~\cite{nashe}, in agreement with~\cite{piee_prev1} and the
predictions of~\cite{prades}.
\begin{table}[h]
\begin{center}
\begin{tabular}{c|c|c|c}
Model&Form-factor&Parameter&Results with Combined error\\
\hline
&$W(z)=G_FM_K^2|f_0|(1+\lambda z)$&$\lambda$&$2.32\pm0.18$\\
Model 1&Linear&$|f_0|$&$0.531\pm0.016$\\
&&$BR_1\times10^{-7}$&$3.05\pm0.10$\\

\hline 
&$W(z)=G_FM_K^2W_+^{pol}+W_+^{\pi\pi}(z)$&$a_+$&$-0.578\pm0.016$\\
Model 2&ChPT at NLO~\cite{piee2}&$b_+$&$-0.779\pm0.066$\\
&&$BR_2\times10^{-7}$&$3.14\pm0.10$\\

\hline 
&$W(\tilde{w},\beta,z)$&$\tilde{w}$&$0.057\pm0.007$\\
Model 3&Combined framework ChPT and &$\beta$&$0.531\pm0.016$\\
&Large-N$_c$ QCD~\cite{piee3}&$BR_3\times10^{-7}$&$3.13\pm0.10$\\
 \hline 
&$ W(z)\equiv W(M_{a},M_{\rho},z) $&$M_{a}$&$0.974\pm0.035$\\
Model 4&The ChPT parametrization involving the &$M_{\rho}$&$0.716\pm0.014$\\
&resonances $a$ and $\rho$ contribution~\cite{piee4}&$BR_4\times10^{-7}$&$3.18\pm0.10$\\

\hline 
n/a &restricted region in $ M_{ee}>140$ MeV/c$^2$&$BR_{mi}\times10^{-7}$&$2.28\pm0.08$\\
\end{tabular}
\hspace*{0.5cm} \caption{Results of fits to the four models, and the
model-independent branching ratio $BR_{mi}(z>0.08).$
\label{tab:resultsrare}}
\end{center}
\end{table}
 A first measurement of the direct CP violating asymmetry of $K^+$
and $K^-$ decay rates in the full kinematic range was obtained by
performing BR measurements separately for $K^+$ and
$K^-$ and neglecting the correlated uncertainties:\\
$\Delta(K^{\pm}_{\pi ee}) = (BR^+-BR^-)/(BR^+ + BR^-) = (-2.2 \pm
1.5_{stat}\pm  0.3_{syst})\%$, where only the uncorrelated
systematic errors have been taken into account. The result is
compatible with no CP violation. However its precision is far from
the theoretical SM expectation~\cite{piee2} of $
|\Delta(K^{\pm}_{\pi ee})|\sim 10^{-5}$ and from the theoretical
SUSY expectation of $ |\Delta(K^{\pm}_{\pi ee})|\sim
10^{-3}$~\cite{cpv_piee_susy}.

\subsection*{$K^\pm \rightarrow \pi^\pm \mu^+ \mu^-$ decays} About
3100 events of the decay $K^\pm \rightarrow \pi^\pm \mu^+ \mu^-$
have been reconstructed from the full data sample. This sample is
four times larger than the existing world statistics~\cite{pdg}. The
main source of background is the abundant $K^\pm\rightarrow3\pi^\pm$
decay (with subsequent $\pi\rightarrow\mu$ decay or with 2
misidentified $\pi$). The background is studied both with data and
MC simulation and is at the order of few percent and well under
control.
\subsection*{Conclusions} In conclusion, precise studies of several
rare kaon decays measurements by the NA48/2 collaboration were
presented. The achieved precisions are similar to or better than the
best previous ones. Due to the simultaneous $K^+$ and $K^-$ beams,
the CP violating charge asymmetry in $K^\pm \rightarrow \pi^\pm e^+
e^-$ was measured for the first time.

\end{document}